\documentclass[preprint, floatfix]{revtex4}

\usepackage[final]{epsfig}
\usepackage{amsmath}
\usepackage{parskip}

\begin{document}

\title{Why the HBT data are not in agreement with a single chemical and kinetic freeze-out}

\author{Thorsten Renk}

\pacs{25.75.-q}
\preprint{DUKE-TH-04-267}

\affiliation{Department of Physics, Duke University, PO Box 90305,  Durham, NC 27708 , USA}

\begin{abstract}

Based on simple scale arguments we argue that any thermalized fireball evolution 
scenario which is in agreement with the observed Hanbury-Brown Twiss (HBT) 
correlation radii at RHIC and reproduces the measured total multiplicity
cannot reflect the distribution of matter at or 
close to a phase transition temperature of $\sim 170$ MeV and any scenario 
which assumes that HBT reflects these properties cannot be reached by a 
sensible choice of evolution parameters while agreeing with all correlation radii.
We investigate this question in more detail using a parametrized version of the
fireball evolution which has been shown to reproduce particle spectra and
HBT for a separate chemical and kinetic freeze-out.

\end{abstract}

\maketitle

\section{Introduction}
\label{sec_introduction}

The measurement of identical two particle correlations in heavy ion collisions is 
capable of providing a wealth of information about the phase space distribution of the emitting
source created in the collision
(for an overview see \cite{HBTReport, HBTBoris}).
It is commonly assumed that the experimentally accessible Hanbury-Brown Twiss correlation functions 
measure properties of the collision system at kinetic decoupling, i.e. the point
of last interaction of the emitted particles. Assuming that the collision
created a thermalized system and taking into account the measured one particle
spectra of different particle species, essential scale parameters of the
system at breakup can be extracted from the data by fitting a suitable parametrized emission
source to both spectra and correlation functions. Examples of such studies are
\cite{FREEZE-OUT} for SPS and \cite{RHIC-HBT, Lisa-BlastWave} for RHIC conditions.

The crucial point in these studies is the following: Single particle spectra or HBT correlations
alone depend on combinations of the temperature $T$ and collective expansion velocity $\overline{v}$ of the
emission source. Therefore, a model fitted to either HBT correlations or spectra
can only determine possible pairs of $(T, \overline{v})$. However, the functional
dependence of single-particle spectra and two-particle correlations on these parameters is different, 
therefore a simultaneous fit to both quantities is able to resolve this ambiguity \cite{FREEZE-OUT}.

Such studies find that kinetic decoupling temperatures less than $120$ MeV are favoured by the data, significantly
below the phase transition temperature $T_C \approx 170$ MeV.
They imply a significant amount of interaction following the phase
transition and consequently hadronic re-scattering cannot be neglected.
On the other hand, statistical models are highly successful in describing the measured
abundancies of hadrons and require hadrochemical freeze-out temperatures very close
to $T_C$ at full SPS and RHIC energies \cite{PBM1,PBM2}. These two observations give rise to the common picture
of separate chemical and kinetic freeze-out.

In contrast, in a series of papers \cite{sf} a single freeze-out model has been
proposed in which both hadron abundancies and momentum spectra
are fixed at the phase transition and there are no significant scattering processes
beyond this point.

A different suggestion has been made in \cite{Wong1, Wong2}. There it was argued
that elastic rescattering processes do not affect the HBT correlation significantly
and therefore the observed HBT parameters would not reflect properties of the
system at kinetic decoupling but at the phase transition or the chemical
freeze-out point where supposedly the last significant inelastic scattering
processes occur.

For SPS conditions there is good experimental evidence for substantial
hadronic rescattering from the dilepton invariant mass spectrum. 
Dynamical models are unable to explain the enhancement of dilepton emission
below the $\rho$ mass seen in the CERES data \cite{CERES} by a possible
quark-gluon plasma (QGP) contribution \cite{Rapp-Dileptons, Synopsis, Amruta, OldDileptons}.
It is difficult to conceive how this region should be filled if 
vacuum properties of hadrons are relevant. In order for a hadronic
contribution to describe the data, either a mass shift of the $\rho$ or
a strong broadening due to its interaction with a hot and dense medium
is necessary.

It is the purpose of this letter to demonstrate that the measured HBT data \cite{Phenix-HBT}
at RHIC cannot
reflect properties of matter near $T_C$ if some simple scale constraints are taken
into account, thus giving evidence that there has to be substantial
inelastic rescattering in the hadronic evolution below $T_C$ 
if one assumes evolution in or close to thermal equilibrium. This result is 
in line with evidence from the simultaneous fits to spectra and HBT correlations 
and the independent evidence from dilepton emission at SPS mentioned above.
It strongly supports the notion that the chemical freeze-out and the kinetic freeze-out
do not occur at the same time.

\section{Some scale arguments}

HBT correlation measurements do not reveal the source geometry as such but rather
measure regions of homogeneity. This can be easily seen in a simple calculation assuming
a Gaussian source: Using this ansatz, for example the relation between the correlation radius
$R_{side}$ and the geometrical (Gaussian) radius of the source $R_G$ is \cite{HBTReport}
\begin{equation}
\label{E-R_side_simple}
R_{side}(m_t) = \frac{R_{G}}{\sqrt{1 + m_t \eta_{\perp f}^2/T_f}}
\end{equation}
where we have introduced the source temperature at kinetic decoupling $T_f$, the transverse mass
$m_t = \sqrt{m^2+ p_\perp^2}$ of the emitted particles and the transverse rapidity $\eta_{\perp f}$
at the Gaussian radius $R_G$ as a measure of the transverse flow 
(assuming a linear increase of the transverse rapidity with radius $r$) at the breakup time.

Expressions assuming other density distributions and flow profiles are naturally different, however the
essential physics is apparent from this particular example: In the limit of vanishing transverse
mass, correlation radius and geometrical radius agree. For finite transverse mass,
the presence of strong flow (large $\eta_\perp$) tends to decrease the correlation radius
by introducing a shift in the average momentum for two particles emitted from
different spacetime points, whereas a large temperature tends to compensate for this effect
by introducing a momentum spread for all particles emitted from a particular point. Thus,
the falloff of the correlation parameter with $m_t$ is governed by the ratio 
$\eta_\perp^2/T_f$. In particular, this implies that if a scenario with $T_f \approx 100$ MeV
can describe the experimental data, the corresponding transverse flow in any
scenario assuming a freeze-out temperature of $170$ MeV has to be stronger
in order to generate the same falloff in transverse mass.

Calculating the total entropy based on the observed multiplicity
and with the help of an equation of state (EOS) based on quasiparticle degrees of freedom
\cite{Quasiparticles} or using directly the EOS as found in lattice
QCD simulations at finite temperature (see e.g. \cite{Lattice}, the difference between
both approaches mainly being the value of the bare quark masses)
we can estimate the volume occupied by hot matter produced
in an Au-Au collision at full RHIC energy at $T_C$
as about $V_{max} \sim 4500$ fm$^3$. This is an upper bound for the volume
of the emission source seen in the HBT data --- any larger volume will
require more entropy than measured experimentally to contain matter at $T_C$.

In the following, we make some very rough estimates for the minimum source volume 
seen in the HBT data. We argue that for $T=T_C$ this volume has to be much larger than the
upper bound $V_{max}$ discussed above. A larger system with the experimentally determined entropy, 
however, must be cooler than 170 MeV, hence the measured HBT correlations cannot originate from
matter at $T = T_C$.

Assuming cylindrical fireball geometry and Gaussian
distributions of matter density in longitudinal and transverse direction, the Gaussian volume at breakup can roughly be
estimated from the correlation radii as $V=(2 \pi)^{3/2} R_{side}^2 R_{long}$.
Here, the correlation radii have to be determined at $m_t = 0$. Experimentally this limit is not accessible,
but we can obtain a lower limit by estimating the volume of the
region of homogeneity for low transverse momenta by using the data in the smallest transverse
momentum bin.

The simple Gaussian ansatz yields $R_{side} \approx 4.9$ fm, $R_{long} \approx 5.9$ fm and a volume of
about 2250 fm$^3$. This seems to be in agreement with the value of $V_{max}$ but misses
out the necessary extrapolation to $m_t = 0$, so the true volume may well
be larger than $V_{max}$. 

In essence, this is a crude estimate, 
but the fact that the volume of the region of
homogeneity is already of the order of magnitude of the maximum possible volume
for matter at $T=T_C$ reveals that there is a constraint
(in the following referred to as 'volume constraint') --- if the flow in the evolution
scenario is such that the region of homogeneity is only about half of the total fireball
volume or less then the extrapolation to the true geometrical size of the emission source
will yield a volume $V > V_{max}$ and the scenario is not compatible with the constraints set by thermodynamics.

There is yet another constraint: Again, for simplicity assuming Gaussian distributions of matter,
the root mean square (rms) radius of the region of homogeneity estimated from
the lowest momentum bin of $R_{side}$ is about 7 fm
(using $R_{rms} = \sqrt{2} R_G$), whereas the
rms radius of the initial gold nucleus (averaged over the centrality bins for which
the HBT correlations are measured) is $\sim 4.6$ fm. Thus, there is an increase
in the rms radius of at least 2.4 fm visible in the data. In a thermal description of the fireball,
transverse flow would be at the origin of the increase in radius and one should expect
$v_\perp^{rms}(\tau) = a_\perp \tau$ and $R_{rms}(\tau) = R_{rms}^0 + a \tau^2/2$ (with
$R_{rms}^0$ the initial rms radius as calculated in overlap calculations and $a_\perp$
the acceleration) to
approximate the expansion. Once the amount of flow is specified
(which can be done by using the one-particle momentum spectra), this yields a solution
$(\tau, a_\perp)$ for the timescale of the transverse expansion. For a moderate flow
of $v =0.5 c$, we find in our rough estimate $\tau = 9.6$ fm/c, for strong 
flow of $v = 0.8c$, we get  $\tau = 6$ fm/c. However, during this time needed for the 
transverse expansion, the longitudinal expansion takes place as well. In a boost invariant scenario,
the longitudinal extension of the system at time $\tau$ (measured in the thermodynamically relevant 
frames locally co-moving with the expanding matter) comes out as
$2 \eta_{max} \tau$. 
The experimentally observed rapidity interval of produced matter is 
$-\eta_{max}\sim -3.5 < \eta < 3.5 \sim \eta_{max}$.
Using the estimate for the time given above and the measured $\eta_{max}$ we find length scales of 
$L = 42 (67)$ fm, well in excess of the
length of the region of homogeneity estimated from the lowest momentum bin of the $R_{long}$ data. 
Thus, during the time necessary for the observed radial expansion, the initial longitudinal
motion of matter in a boost-invariant scenario leads to a large longitudinal extension
which in turn implies a violation of the volume constraint (we refer to this as 'time constraint' in the following).

Since the rapidity distribution of matter is a measured quantity, this constraint cannot
be avoided by a slower boost-invariant expansion. However, longitudinal stopping and re-acceleration 
as in the Landau scenario can lead to a significantly smaller longitudinal extension 
for a given expansion time $\tau$ while the same rapidity distribution can be achieved.
Thus a Bjorken expansion picture cannot possibly be reconciled with the assumption of a freeze-out at 
the phase transition; at least some degree of a Landau-type initial longitudinal
compression and re-expansion of matter must be assumed in order to match the scales.

\section{A detailed investigation}

In this section, we will investigate the rough estimates of the previous section using a
model framework for the fireball expansion which in essence is a parametrization
inspired by a hydrodynamical evolution of the collision system. This model has been
shown to give a good description of both single particle spectra and two particle
correlations at RHIC simultaneously for a breakup temperature well below the
phase transition temperature. It is described
in greater detail in \cite{RHIC-HBT}, here we only repeat the essential facts:

For the entropy density at a
given proper time we make the ansatz 
\begin{equation}
s(\tau, \eta_s, r) = N R(r,\tau) \cdot H(\eta_s, \tau)
\end{equation}
with $\tau $ the proper time measured in a frame co-moving with a given volume element, 
$\eta_s = \frac{1}{2}\ln (\frac{t+z}{t-z})$ the spacetime
rapidity and $R(r, \tau), H(\eta_s, \tau)$ two functions describing the shape of the distribution
and $N$ a normalization factor. We use Woods-Saxon distributions 
\begin{equation}
\begin{split}
&R(r, \tau) = 1/\left(1 + \exp\left[\frac{r - R_c(\tau)}{d_{\text{ws}}}\right]\right)
\\ & 
H(\eta_s, \tau) = 1/\left(1 + \exp\left[\frac{\eta_s - H_c(\tau)}{\eta_{\text{ws}}}\right]\right).
\end{split}
\end{equation}
for the shapes. Thus, the ingredients of the model are the skin thickness 
parameters $d_{\text{ws}}$ and $\eta_{\text{ws}}$
and the para\-me\-tri\-zations of the expansion of 
the spatial extensions $R_c(\tau), H_c(\tau)$ 
as a function of proper time. From the distribution of entropy density, the thermodynamics can be inferred
via the EoS and particle emission is then calculated using the Cooper-Frye formula.
In \cite{RHIC-HBT}, the model parameters have been adjusted such that the model gives a good
description of the data.

It is the aim of the present letter to test how well a description of the HBT correlation measurements
is possible if one assumes that the HBT measurement reflects the properties of matter at
the phase transition, either because of a common chemical and thermal freeze-out of
because elastic rescattering does not modify the correlations. The strategy is the
following: Using a set of assumptions characterizing a given scenario, we will tune the remaining
parameters such that a good description of one HBT parameter is achieved and investigate how well
the other correlation radii are reproduced by this choice.

For simplicity, we take the transverse acceleration $a_\perp$ to be a constant (i.e. independent
of the EoS) and only fix the final value $v_\perp^{rms}$ at the rms radius of the fireball. 

\subsection{Strong longitudinal constraints}

\begin{figure*}[!htb]
\begin{center}
\epsfig{file=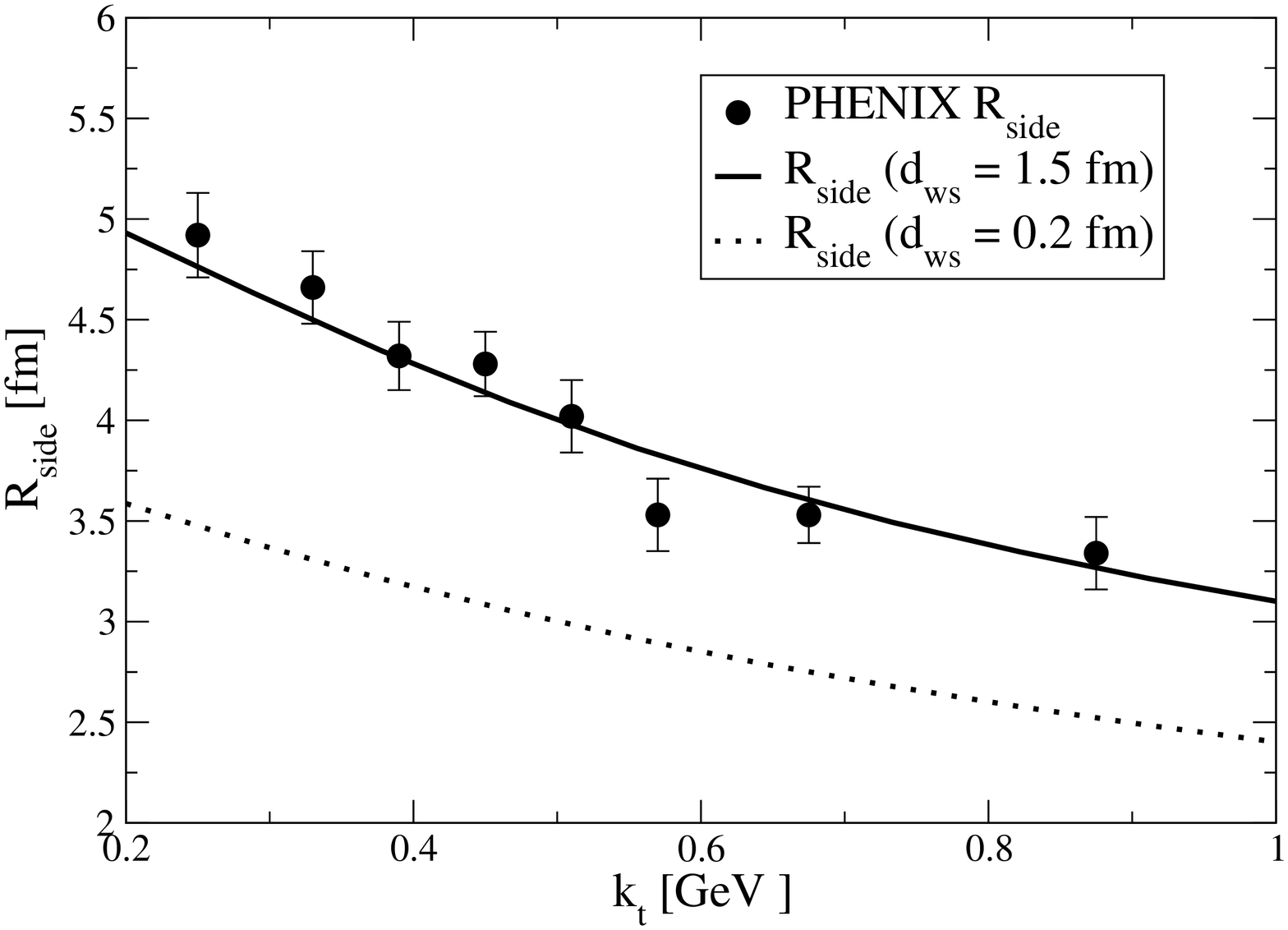, width = 6.3cm, angle=-90}
\epsfig{file=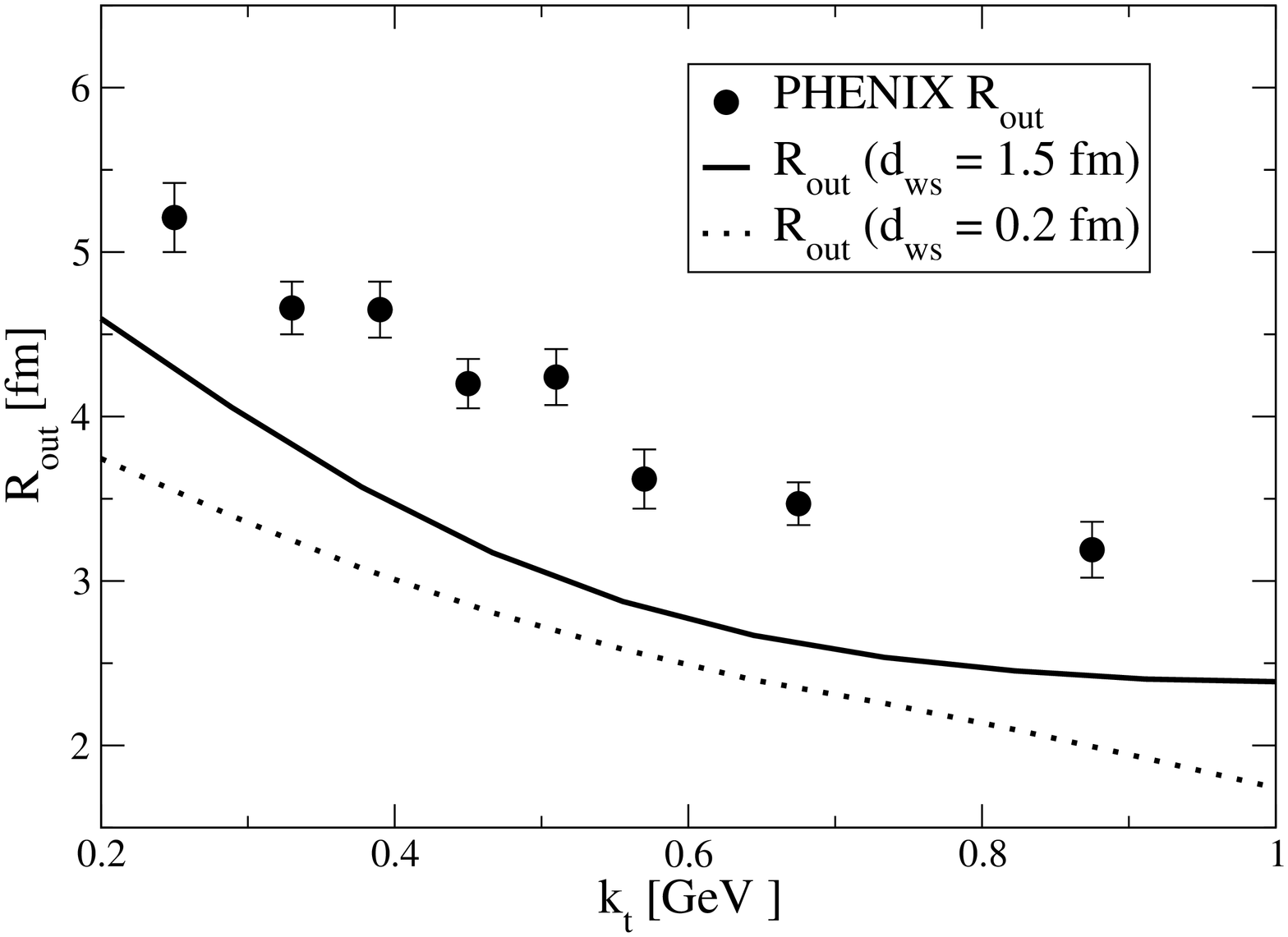, width=6.3cm, angle=-90}
\epsfig{file=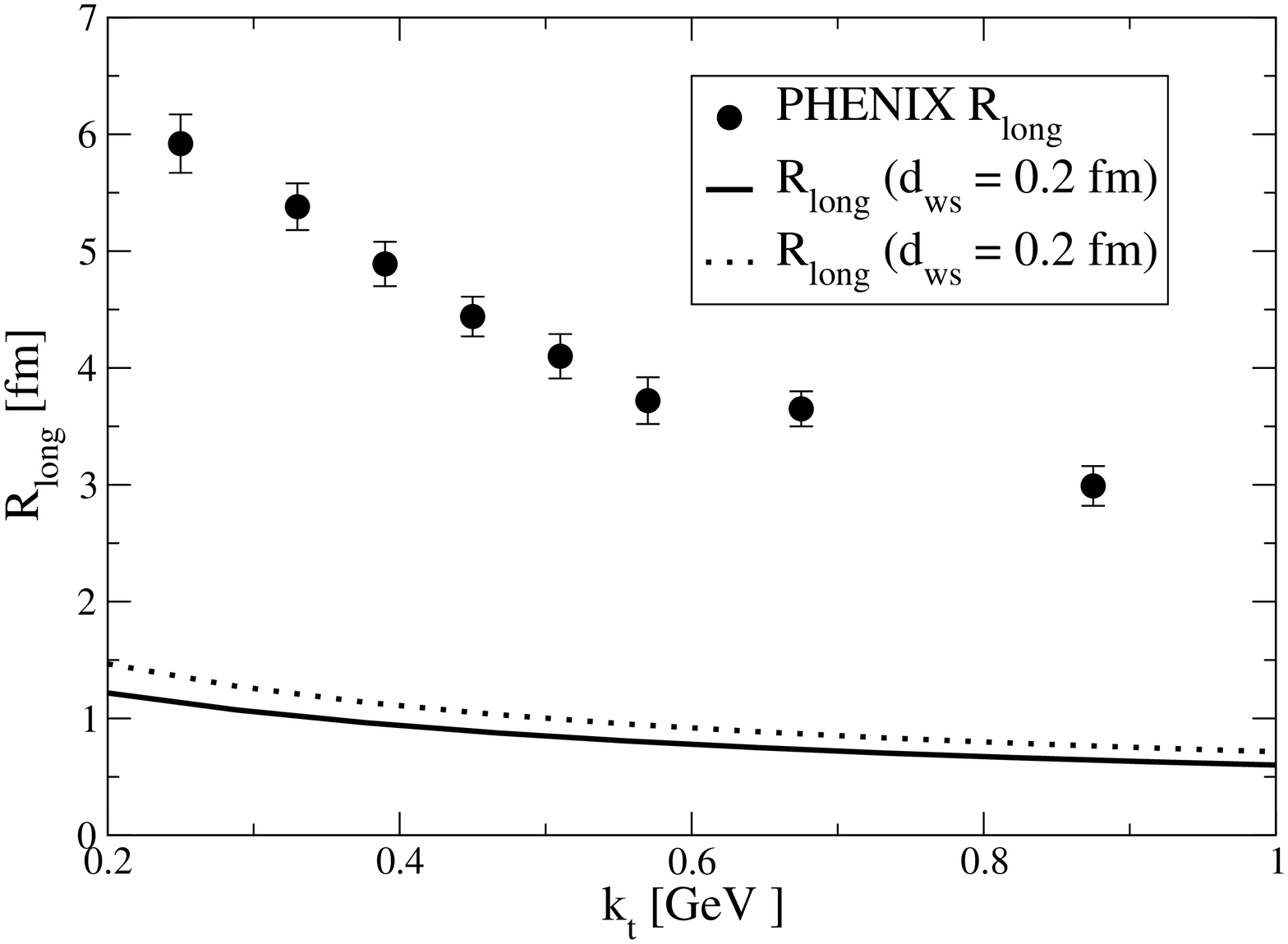, width = 6.3cm, angle=-90}
\epsfig{file=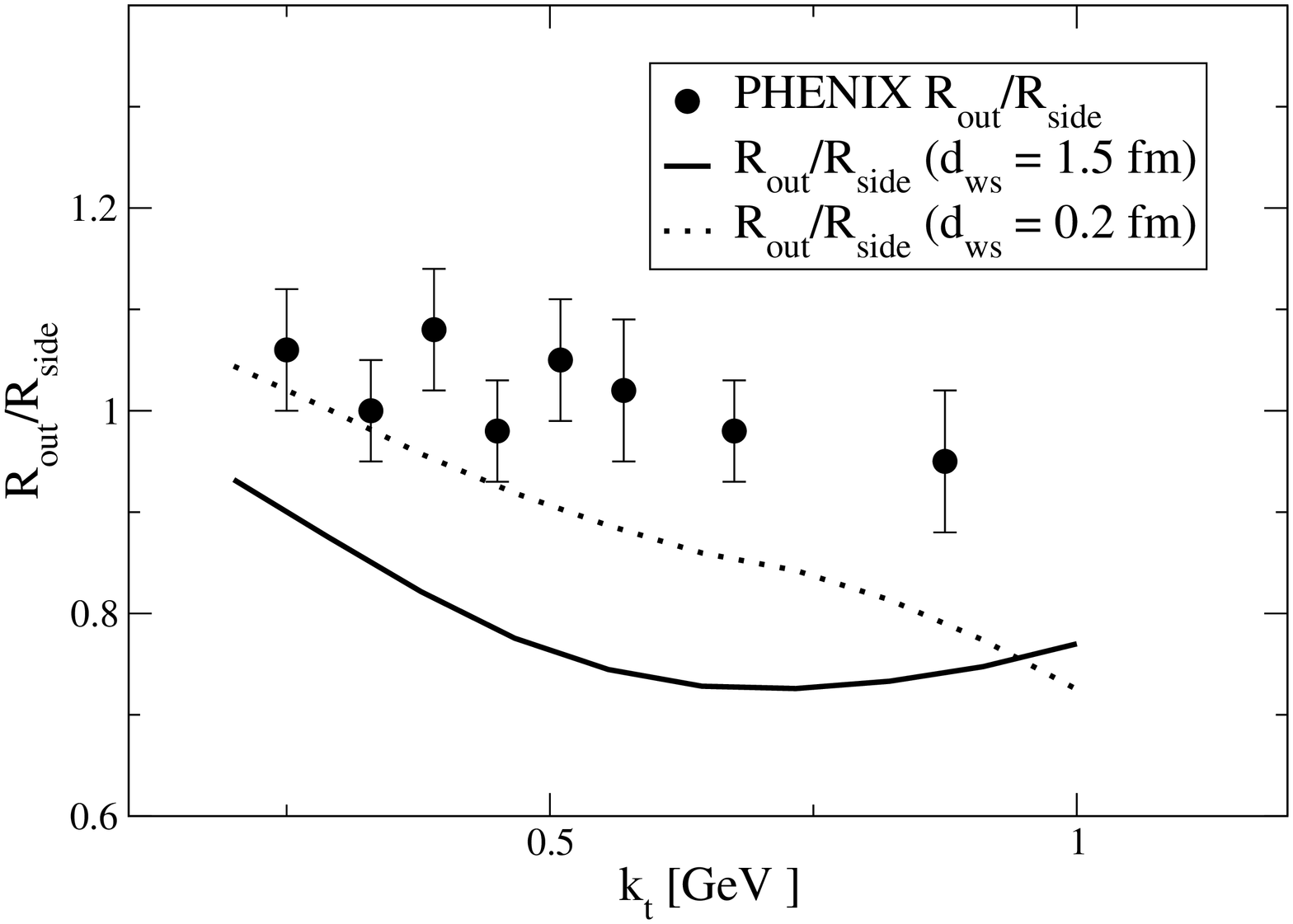, width=6.3cm, angle=-90}
\end{center}
\caption{\label{F-FullStop}The HBT correlation parameters $R_{\text{side}}, R_{\text{out}}, R_{\text{long}}$
and the ratio $R_{\text{out}}/R_{\text{side}}$ in the model calculation as compared to PHENIX
data \cite{Phenix-HBT}, assuming almost complete initial compression of matter. Shown are the
results corresponding to two different skin thickness parameters $d_{ws}$ of the entropy density
distribution.}
\end{figure*}

In a first calculation, we assume that the fireball evolution incorporates a simultaneous
chemical and kinetic freeze-out at the phase transition. This implies that the rapidity
interval filled by the matter at breakup has to agree with the experimentally observed
rapidity interval, i.e. the emission source has to fill about 7 units of rapidity.

Assuming a Bjorken boost-invarinat non-accelerated expansion scenario, this strong
longitudinal flow implies a quick volume expansion. Thus, the volume constraint
is hit early before transverse flow could expand the system and no satisfactory
description of $R_{side}$ is possible since the geometrical radius ends up being
smaller than the radius of the homogeneity region.

However, going to the opposite limit of almost completely stopped matter ($-0.2 <\eta_0<0.2$)
and accelerated re-expansion to the finally observed interval a good description of
$R_{side}$ is possible with a large surface thickness parameter $d_{ws} \sim 1.5$ fm and strong
transverse flow  $v_\perp^{rms} \approx 0.8$. This, however, is not caused by a dynamical
expansion but by the fact that the highly compressed initial state in combination 
with the surface smearing leads to a huge initial temperature and a Cooper-Frye surface far
away from the rms radius of the entropy density distribution. In fact, the hypersurface
moves inward as the longitudinal acceleration leads to rapid cooling of the system in
the later stages.

The resulting correlation radii are shown (for comparison also using a sharp surface with
$d_{ws} = 0.2$ fm) in Fig.~\ref{F-FullStop}. The largest discrepancy is seen
in $R_{long}$. It is evident that the system did not have
enough time for longitudinal expansion such that the normalization of $R_{long}$ could
be reproduced. The falloff in $m_t$ is however in agreement with the data (as can be seen
by rescaling the curve with a constant factor) as it should since the scenario reproduces
the experimentally observed rapidity interval. The normalization is off by a factor of
$\sim 4$ which cannot easily be accounted for by a peculiarity of the present model.

\subsection{Weak longitudinal constraints}

Assuming that there are still elastic rescattering processes going after the phase transition,
there's no reason why the rapidity interval filled by matter at the phase transition point
should agree with the experimentally accessible rapidity interval which reflects properties
of matter at the later kinetic freeze-out. If the HBT correlation radii would be unaffected by
elastic rescatterings, the rapidity interval relevant for $R_{long}$ would be that at 
the phase transition.

In principle, this would allow for longitudinal expansion slower than in the previous
case, leading to more transverse expansion and improving the agreement with the measured
HBT radii. However, the falloff of $R_{long}$ with $m_t$ demands a longitudinal velocity gradient
which is compatible with the measured rapidity interval. The only longitudinal ambiguity
concerns the amount of initial longitudinal compression, and the relevant parameter
$\eta_0$ can be fit to the absolute magnitude of $R_{long}$.

\begin{figure*}[!htb]
\begin{center}
\epsfig{file=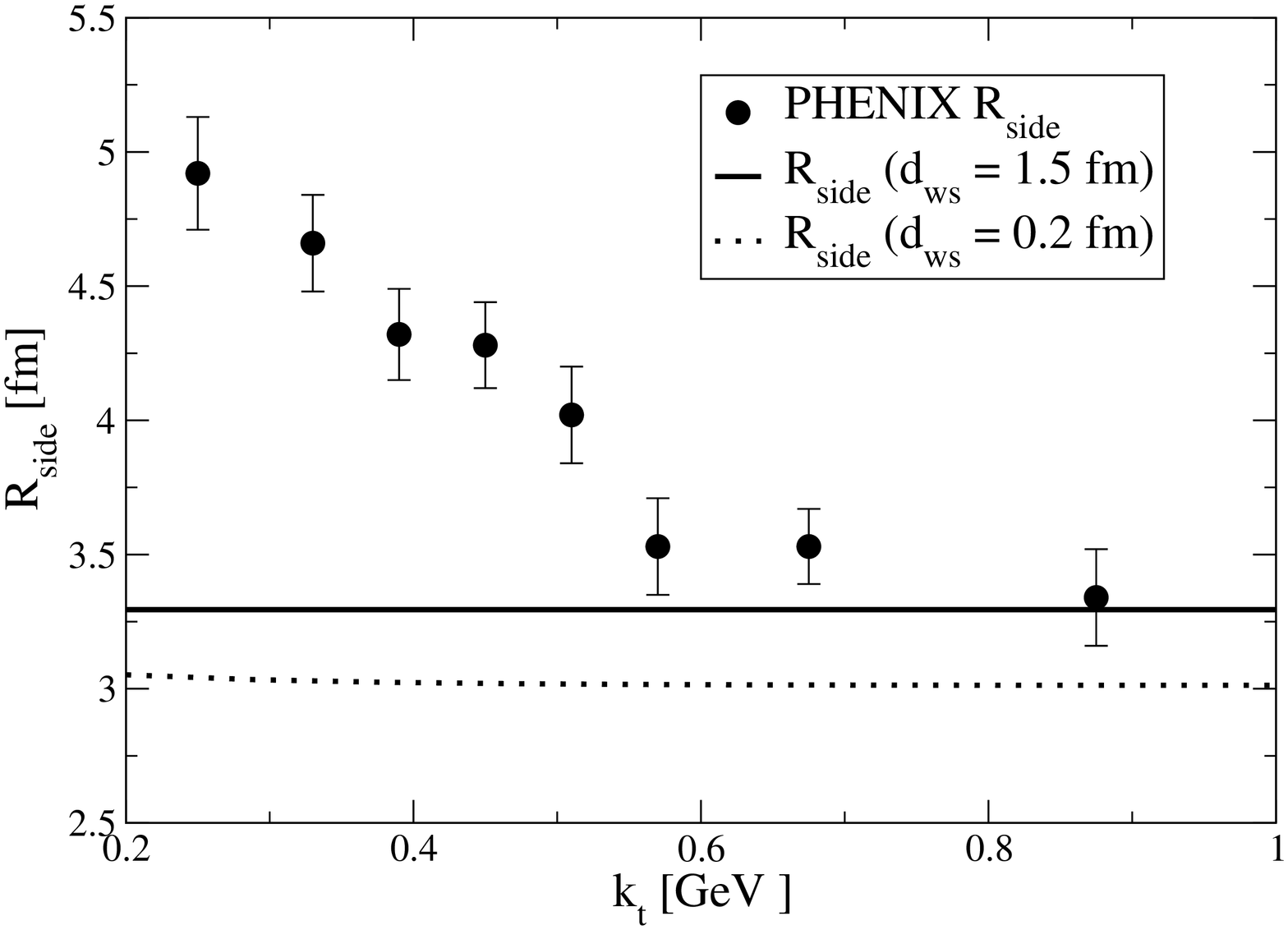, width = 6.3cm, angle=-90}
\epsfig{file=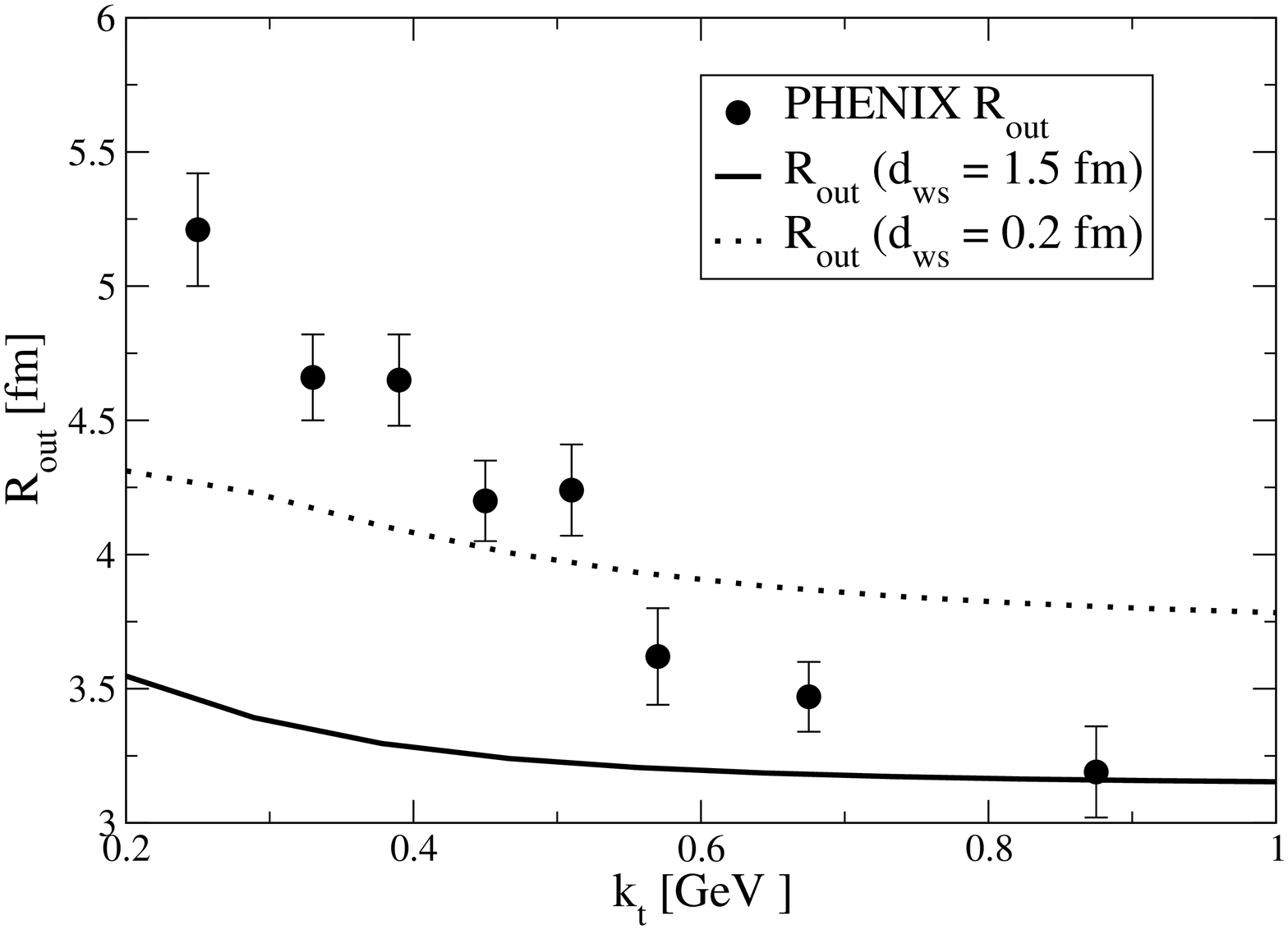, width=6.3cm, angle=-90}
\epsfig{file=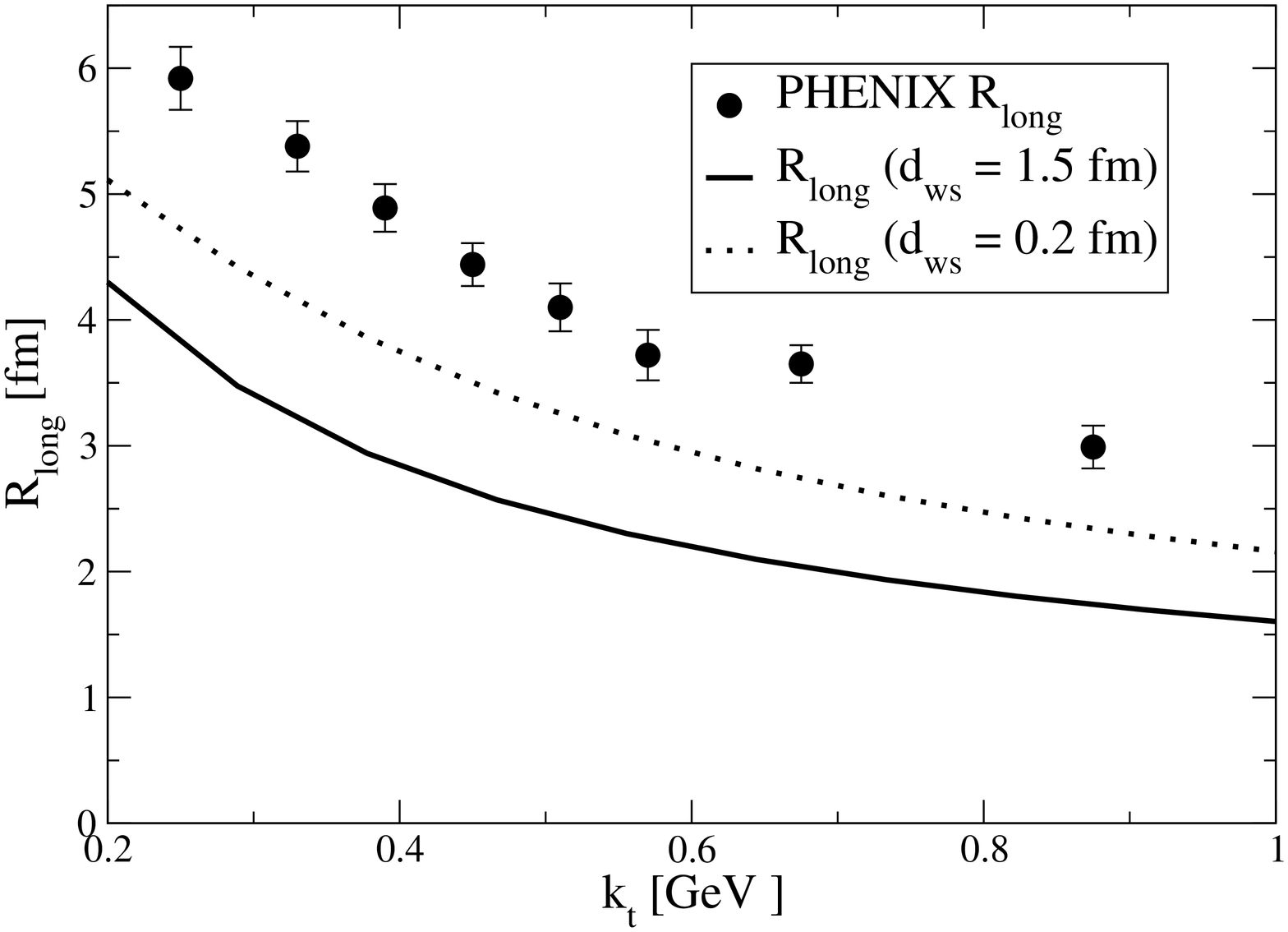, width = 6.3cm, angle=-90}
\epsfig{file=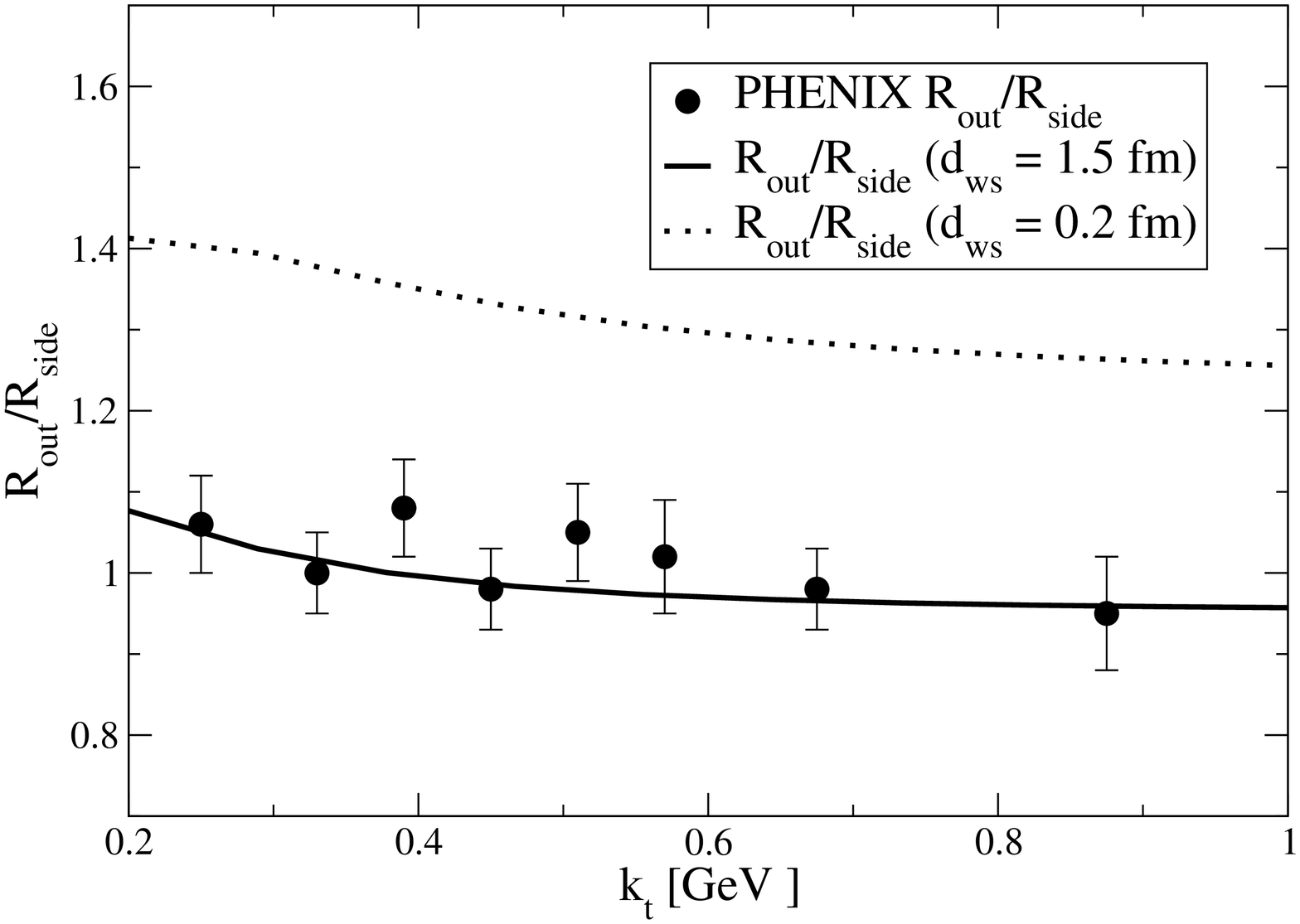, width=6.3cm, angle=-90}
\end{center}
\caption{\label{F-R_long}The HBT correlation parameters $R_{\text{side}}, R_{\text{out}}, R_{\text{long}}$
and the ratio $R_{\text{out}}/R_{\text{side}}$ in the model calculation as compared to PHENIX
data \cite{Phenix-HBT}, aiming for the best possible description of $R_{long}$ assuming
vanishing transverse flow. Shown is the result corresponding to two different skin thickness
parameters $d_{ws}$ of the entropy density distribution.}
\end{figure*}

In doing so, however, it turns out that the absolute magnitude of $R_{long}$ cannot
be fitted even for vanishing transverse flow - the longitudinal expansion 
with the observed velocity gradient alone requires a volume which is not in agreement
with the volume constraint --- in spite of the correct falloff, the resulting
curve is 20\% below the data (and, having no transverse flow, the resulting
scenario fits neither $R_{side}$ nor the transverse mass spectra).
This is  apparent from Fig.~\ref{F-R_long}.

The result is readily interpreted: A sharp surface of the radial density distribution leads to
a stationary Cooper-Frye surface and hence to less emission before breakup than a dilute distribution which
implies a receding Cooper-Frye surface. Thus, the system has a slightly longer lifetime for
$d_{ws} = 0.2$ fm/c, leading to more expansion and a better description of $R_{long}$. On the other
hand, a dilute distribution initially implies a Cooper-Frye surface farther out and hence a slightly
larger $R_{side}$. The behaviour of $R_{side}$ and $R_{long}$ illustrates nicely the tradeoff
between longitudinal and transverse extension caused by the volume constraint.

\section{Summary}

We have shown both by simple scale arguments and in more detailed studies that a freeze-out
temperature of 170 MeV cannot be compatible with the measured data. Thus, the 2-particle
correlations do not give any indication that there would be either a simultaneous chemical
and kinetic freeze-out or that only elastic scattering processes which do not
modify the HBT correlations prevail after the phase
transition.

This is in essence due to the fact that the falloff of the HBT radii with
transverse momentum requires a strong flow, even more so for a comparatively large
temperature of 170 MeV. In the presence of strong flow however the region of homogeneity
seen in the correlation parameters is smaller than the actual volume, hence the true
geometrical volume necessary for these radii exceeds by far the volume determined by
the EoS and the total fireball entropy.

This is not an artefact of the model - mismatches between calculation and data
for the best choice of parameters of order 3-4 indicate that this is a fundamental
problem. We have not even made an attempt to simultaneously
describe single particle spectra for these emission temperature and flow combinations.

The logical conclusion is then that chemical and thermal freeze-out are separate phenomena
and that the HBT correlation radii reflect indeed properties of matter at or close to the
kinetic freeze-out temperature. This requires frequent interactions in the hadronic
evolution phase and the production of resonance states, a conclusion which
is directly confirmed in the SPS case by the measured dilepton invariant mass spectrum.


\begin{acknowledgments}

I would like to thank S.~A.~Bass and B.~M\"{u}ller for helpful discussions, comments and their
support during the preparation of this paper.
This work was supported by the DOE grant DE-FG02-96ER40945 and a Feodor
Lynen Fellowship of the Alexander von Humboldt Foundation.
\end{acknowledgments}

\end{document}